\pdfoutput=1
\documentclass[epj]{svjour}
\usepackage{graphicx,amsfonts,amssymb}

\newcommand{\middlefig}{.45\textwidth}
\newcommand{\smallerfig}{.23\textwidth}

\begin{document}

\title{\bf Dark lattice solitons in one-dimensional waveguide arrays with defocusing saturable nonlinearity and alternating couplings}
\titlerunning{Discrete dark solitons in
waveguide arrays with saturable nonlinearity and alternate couplings}

\author{Andrey Kanshu\inst{1} \and Christian E. R\"uter \inst{1} \and Detlef Kip\inst{1} \and Jes\'us Cuevas\inst{2} \and Panayotis G. Kevrekidis\inst{3}}
\institute{Faculty of Electrical Engineering, Helmut Schmidt University, 22043 Hamburg, Germany \and %
Grupo de F\'{\i}sica No Lineal, Departamento de F\'{\i}sica Aplicada I, Universidad de Sevilla.
Escuela Polit\'ecnica Superior, C/ Virgen de \'{A}frica, 7, 41011
Sevilla, Spain \and %
Department of Mathematics and Statistics, University of Massachusetts,
Amherst MA 01003-4515, USA
}

\date{\today}

\abstract{
In the present work, we examine ``binary'' waveguide arrays, where
the coupling between adjacent sites alternates between two distinct
values $C_1$ and $C_2$ and a saturable nonlinearity is present on
each site. Motivated by experimental investigations of this type
of system in fabricated LiNbO$_3$ arrays, we proceed to analyze
the nonlinear wave excitations arising in the self-defocusing
nonlinear regime, examining, in particular, dark solitons
and bubbles. We find that such solutions may, in fact, possess
a reasonably wide, experimentally relevant parametric interval
of stability, while they may also feature both prototypical
types of instabilities, namely exponential and oscillatory ones,
for the same configuration. The dynamical manifestation of the instabilities
is also examined through direct numerical simulations. }

\maketitle

\section{Introduction}

Dark solitary waves are ubiquitous nonlinear excitations of
dispersive wave models with a so-called self-defocusing nonlinearity.
Perhaps the prototypical model where they arise is the nonlinear
Schr{\"o}dinger equation with a defocusing nonlinearity~\cite{kivshar},
as it is referred to in nonlinear optics (since it induces the
spreading of optical beams). These excitations consist of a density
dip (i.e., a dark notch) accompanied by a phase jump across the
density minimum. In addition to their observation in nonlinear optics
extensively summarized in~\cite{kivshar,kivsharbook}, numerous experiments
have demonstrated the emergence of these nonlinear states
 in the atomic physics
of Bose-Einstein condensates, as has recently been comprehensively
reviewed in~\cite{djf}. Interestingly, the physical relevance of dark
solitons are not limited to these areas but rather also extend
to parametrically-driven shallow liquids~\cite{djf5}, discrete
mechanical systems~\cite{djf6}, electrical lattices with
nonlinear capacitors~\cite{lars}, thin magnetic films~\cite{djf7} and
dissipative variants thereof in complex plasmas~\cite{djf8}, among others.

In recent years, there has been a considerable interest in the
realization of such excitations in systems bearing lattice potentials,
to appreciate the effects of discreteness on the existence and
stability, as well as on the dynamics and mobility
of such excitations; see e.g. the prototypical theoretical
study of \cite{johkiv} and even the earlier work of \cite{chuby}.
 Experimental developments have enabled the
realization of such excitations, especially so in arrays of
optical waveguides either in the context of AlGaAs in the
anomalous diffraction regime with the cubic nonlinearity due
to the Kerr effect~\cite{silb}, or in defocusing lithium
niobate waveguide arrays, which exhibit a different type of nonlinearity,
namely a saturable, defocusing one due to the photovoltaic effect \cite{kip}.
These experimental investigations, in turn, led to comparative
studies between the features of the dark solitons in these two
discrete models~\cite{fitrakis,dark},
as well as to the examination
of states consisting of multiple dark solitary waves~\cite{susanto,pelinovsky}.
These findings have been summarized in a recent book~\cite{pgk}.
Furthermore, in these waveguide array systems, not only have dark
solitons been identified in higher gaps
(of the associated periodic potentials)~\cite{rongdong1},
but they have also been found
to arise as members of multi-component soliton complexes
such as dark-bright solitary waves~\cite{rongdong2}.

In the present work, we consider a nontrivial variant of the
above nonlinear dynamical lattices. In particular, we focus on
the setting of ``heterogeneous'' lattices and more specifically
in the context of lattices with alternating couplings between
two distinct values $C_1$ and $C_2$.  The existence and
stability of bright solitons in binary lattices has been
investigated both theoretically~\cite{Sukhoroukov2002,Vicencio2009}
and experimentally~\cite{Morandotti2004,Kanshu2012}. It should be added,
however, that binary lattices can be constructed in different ways,
such as e.g. by waveguides of the same separation between the channels,
but alternating in widths as in~\cite{Morandotti2004}, or by ones
of the same width but alternating in separation as in~\cite{Kanshu2012}.
Here, in line with the latter approach,
we fabricate one-dimensional arrays of such
lattices in a nominally non-doped LiNbO$_3$ substrate
by Ti in-diffusion.
In this case the underlying
linear spectrum possesses two pass bands, with a mini-gap
between them (as opposed to the single band of the homogeneous
waveguide array).
In the gaps of the linear spectrum, we will seek nonlinear wave
solutions, such as dark solitons and bubbles in what follows.
We find that both types of
obtained nonlinear waves
may possess both regions of stability as well as
regions of both types of instability, exponential as well as oscillatory,
as the propagation constant parameter is varied. This
is contrary to what is the case
for the homogeneous (coupling)
counterpart of the model, where only oscillatory
(or only exponential) instabilities may be observed for a particular
type of configuration. When the waveforms our found to be unstable,
direct numerical simulations of the system are used to explore the
evolution of the instability, giving rise to the
breakup of the stationary structure into moving states or to its breathing
dynamics.

Our presentation is structured as follows. In section II, we present
the experimental setup and some prototypical experimental results.
These, in turn, motivate the theoretical model presentation
and discussion of section III and
the systematic numerical investigation of section IV. Finally, in section
V, we summarize our findings and propose some potential directions for
future study.

\section{Experimental Motivation}

\begin{figure}[!ht]
\begin{center}
    \includegraphics[width=\middlefig]{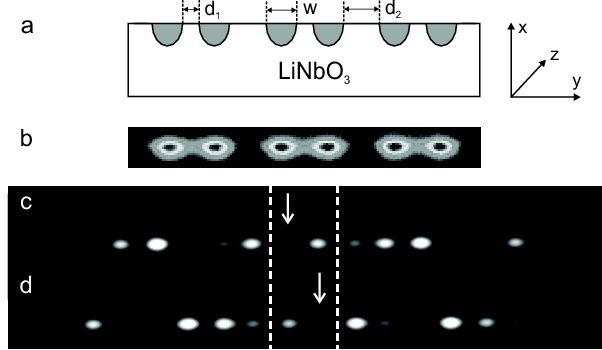}
\caption{Geometry of the binary waveguide array (a), experimental output intensity distribution for homogeneous array excitation with Bloch momentum $k=0$ (b), output intensity distribution when a single element of the binary lattice is excited (see arrows), either on the left (c) or the right hand side (d). }
\label{fig:exp1}
\end{center}
\end{figure}

For our experiments, we fabricate a one-dimensional waveguide array in a nominally non-doped LiNbO$_3$ substrate. Using standard photolithographic techniques, we form, on the substrate's surface, an array with 250 Ti stripes of equal width $W=4.0\,\mu$m and alternating distances $d_{1}=2.5\,\mu$m and $d_{2}=4.5\,\mu$m between them. By in-diffusion of Ti stripes at $T=1000\,^{o}$C and following polishing, we obtain samples with dimensions of 1\,mm$\,\times\,20\,$mm$\,\times\,7.8\,$mm, where the ferroelectric c-axis is pointing along the 7.8\,mm-long direction, and channels are aligned along the 20\,mm-long propagation direction [Fig. \ref{fig:exp1}(a)]. By choosing an appropriate thickness of the sputtered Ti stripes, and duration and temperature of in-diffusion, we form a sample possessing only a single allowed band [Fig. \ref{fig:exp1}(b)-(d)]. This allows us to use standard end-facet coupling for studying linear and nonlinear light propagation in the lattice. Contrary to the single uniform lattice, the first band of our sample is split into two parts, separated by an additional mini-gap that opens in the middle of the Brillouin zone. Finally, this sample possesses a defocusing saturable nonlinearity which is probably due to small impurity concentrations incorporated into the substrate during high-temperature treatment, as well as
intrinsic photorefractive defects like Nb on Li sites in congruently melting LiNbO$_3$. At slightly higher intensities or input powers [of the order of (10-100)$\,\mu$W per channel] the observed nonlinear index changes of this sample are of approximately the same magnitude when compared to, for example, those fabricated on Fe-doped substrates.

To form dark solitons in our sample, we focus an appropriate light amplitude distribution onto the polished input facet, and record the temporal evolution of the out-coupled light by means of a 20$\times$ microscope objective and a CCD camera connected to a computer. With the help of an additional plane reference wave we are able to monitor the phase distribution of the out-coupled light, too. For the case of a binary lattice different kinds of dark soliton solutions exist: a fundamental dark soliton in the semi-infinite gap, a bubble-like dark soliton that is located in the semi-infinite gap, too, and another dark soliton existing in the extra mini-gap having an alternating phase distribution. All solutions require excitation using a suitable amplitude profile and appropriate power to be formed. Because the nonlinearity of our sample is of saturable nature, we can, in a certain range, vary the input light power accordingly in order to generate the necessary nonlinear index changes for dark soliton formation.

\begin{figure}[!ht]
\begin{center}
    \includegraphics[width=\middlefig]{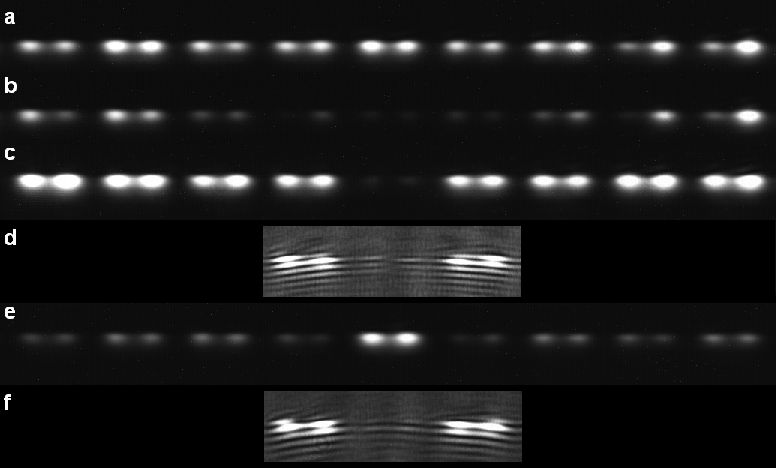}
\caption{Light distribution on the output surface of the lattice by homogeneous excitation on input surface (a), discrete diffraction (b), dark soliton (c), interferogram of the dark soliton showing the $\pi$ phase jump (d), guiding properties of a dark soliton when the lattice is excited with a homogeneous probe beam (e), and interferogram of a bubble soliton when both sides of the dark notch are in phase (f).}
\label{fig:exp2}
\end{center}
\end{figure}

For the formation of a dark soliton in the semi-infinite gap above the center of the Brillouin zone at $k=0$, we use a broad input beam with homogeneous intensity distribution that is focused into the lattice with the help of a cylindrical lens. This beam passes a phase mask which imparts an additional phase of $\pi$ on half of the beam. The dark notch generated in this way is adjusted to coincide with the center of one element of the binary lattice. First, in Fig. \ref{fig:exp2}(a) we show the output intensity distribution, when the lattice is excited on the input surface homogeneously without using the phase mask. Adding the phase mask into the beam and using low input powers of the order of nW, we observe discrete diffraction of the generated dark notch in Fig. \ref{fig:exp2}(b). With increasing power, the sample starts to behave nonlinearly and the width of the dark notch decreases with increasing recording time. After a few minutes it evolves finally into a dark soliton shown in Fig. \ref{fig:exp2}(c), where the width of the dark notch matches the one on the input facet. This nonlinear state is robust and can be observed for more than one hour without noticeable changes. The corresponding phase relation is demonstrated in the interferogram presented in Fig. \ref{fig:exp2}(d), which clearly shows the phase jump of $\pi$ in the center of the structure. The generation of a dark soliton leads to a change of the corresponding refractive index profile and the formation of a positive defect located at the position of the dark notch. We test the guiding properties of this defect by launching a broad probe beam of low intensity having a plane phase front, and observe trapping of light inside the recorded defect in Fig. \ref{fig:exp2}(e). The experimental setup used in this work is similar to the one in Ref.~\cite{kip}, but with the two-beam interference on the input side replaced by a phase mask. For the formation of a dark soliton in the semi-infinite gap above the center of the Brillouin zone at $k=0$, we use a broad input beam (diameter 6\,mm) of wavelength 532\,nm with homogeneous intensity distribution that is focused into the lattice with the help of a cylindrical lens ($f=100$\,mm). This beam passes a phase mask (half-wave MgF$_{2}$ layer evaporated on glass substrate) which
imparts an additional phase of $\pi$ on half of the beam.

When the phase mask used to add the extra phase jump of  $\pi$ on the input light distribution is misadjusted (tilted), a dark notch (generated in part by a shadow region induced by the mask) with both sides of the structure being in phase is formed. Using this input light distribution on the input facet, we are able to experimentally observe another nonlinear localized mode: The generated bubble solution has a similar intensity distribution as the dark soliton [Fig. \ref{fig:exp2}(c)], but without a phase jump of $\pi$ in the center. The corresponding phase distribution on the output facet is given in Fig. \ref{fig:exp2}(f). Note that in the experiment this nonlinear state is found to be unstable for long recording time, but may be observed during the initial time of recording only. For longer recording or when the input power is further increased, a transformation to the dark soliton solution described previously occurs. This experimental
observation suggests that the dark soliton should be structurally more
robust than the bubble waveform.

To form a dark soliton in the mini-gap one needs to excite each dimer of the lattice in phase, with alternating phase jumps between neighboring units.
Additionaly, a $\pi$ phase jump inside the central element is required to generate a dark notch. By using an interference pattern with a modulation adjusted to fit to the grating period of our lattice, and by passing this beam through an additional phase mask for the generation of the dark notch, a suitable input
light distribution has been generated. However, due to limited precision in
the adjustment (this was mainly the limited symmetry in exciting each dimer
with exactly equal powers in both channels), no clear discrete diffraction
pattern was observed for this situation on the output. Thus, we were unable
to perform the corresponding nonlinear experiments. A possible solution might
be the use of a spatial light modulator to create a more precisely
defined input profile, which was not available presently in the experimental
setup used. It should be noted here that these ``staggered'' states have,
in fact, been accessible to our numerical investigations. Nevertheless, the
lack of experimental control over their generation indicated above, as well
as equally importantly, our generic observation of their large scale
instability and the rapid evolution of their unstable background
towards lattice dynamical turbulence
have precluded us from considering these states further herein.

\section{Theoretical Model and Analytical Considerations}
In order to theoretically model a setup corresponding to our experimental
configuration, we consider a set of discrete nonlinear Schr\"odinger
(DNLS) equations with a saturable onsite nonlinearity (see also~\cite{kip})
in the form:

\begin{eqnarray}
    i\frac{dA_j}{dz}+(C_1B_j+C_2B_{j-1})+\alpha\frac{|A_j|^2}{1+\kappa|A_j|^2}A_j &=& 0, \\
    i\frac{dB_j}{dz}+(C_1A_j+C_2A_{j+1})+\alpha\frac{|B_j|^2}{1+\kappa|B_j|^2}B_j &=& 0
\end{eqnarray}
The defocusing nature of the nonlinearity is encapsulated in
the value of $\alpha=-1$; furthermore, we have used a value of
$\kappa=5\times10^{-4}$ which has been found to accurately
account for the saturation effect.
The electric field of interest $E_n$ is defined from $A_j$ and $B_j$ as:
\begin{equation}
    E_{2j}=A_j,\qquad E_{2j+1}=B_j.
\end{equation}

Within this model, we seek stationary solutions in the form:
\begin{equation}\label{eq:stat}
    A_n=\exp(-i\beta z)a_n,\qquad B_n=\exp(-i\beta z)b_n.
\end{equation}
For the experimental waveguide array, by matching information about the
linear diffraction properties of the lattice, we have inferred that
$C_1=0.43$ mm$^{-1}$ and $C_2=0.14$ mm$^{-1}$ provide reasonable
approximations to the experimental diffraction pattern
observations [see Fig.~1(c,d)]. In light of that,
we have fixed the value of $C_1$ to $0.43$ mm$^{-1}$, but in order
to appreciate the effect of variation of the spacing, we have looked
at the relevant phenomenology as a function of the parameter $C_2$.
Furthermore, we have varied the propagation constant $\beta$, in order
to explore the family of nonlinear stationary, potentially observable
solutions.

We have focused, in particular, on two different types of solutions,
namely dark~\cite{susanto,dark} and bubble~\cite{bubble} solitons, chiefly with a uniform
background
and ${a_n,b_n}\in\mathbb{R}\,\forall n$.

Upon identifying the relevant solutions through a fixed
point computation, we have proceeded to examine
their
spectral stability properties by considering a linearization analysis.
To that effect, small perturbations [of order ${\rm O}(\delta)$,
with $0< \delta \ll 1$] are introduced in the form

\begin{eqnarray}
    && A_n(z,x)=e^{-i \beta z} \left[A_{n,0} + \delta (P_n e^{i \omega z} + Q_n^{*} e^{-i \omega^{*} z}) \right]~, \nonumber \\
    && B_n(z,x)=e^{-i \beta z} \left[B_{n,0} + \delta (R_n e^{i \omega z} + S_n^{*} e^{-i \omega^{*} z}) \right]~,
\end{eqnarray}
and the ensuing linearized equations are then solved to O$(\delta)$, leading
to the following eigenvalue problem:
\begin{equation}\label{eq:stability}
    \omega
    \left(\begin{array}{c} P_n \\ Q_n \\ R_n \\ S_n \end{array}\right)=
    \left(\begin{array}{cccc}
        L_1(a_{n,0}) & L_2(a_{n,0}) & H^{-}_n & 0 \\ -L_2(a_{n,0}) & -L_1(a_{n,0}) & 0 & -H^{-}_n \\
        H^{+}_n & 0 & L_1(b_{n,0}) & L_2(b_{n,0}) \\ 0 & -H^{+}_n & -L_2(b_{n,0}) & -L_1(b_{n,0})
    \end{array}\right)
    \left(\begin{array}{c} P_n \\ Q_n \\ R_n \\ S_n \end{array}\right),
\end{equation}

for the eigenfrequency $\omega$ and the associated eigenvector $(P_n,Q_n,R_n,S_n)^T$, where
$L_1,\,L_2,\,H^{\pm}_n$ are the following operators:

\begin{eqnarray}
    L_1(x) &=& \beta+\nu\frac{2|x|^2+|x|^4}{(1+|x|^2)^2}~, \nonumber
\\[1.0ex]
    L_2(x) &=& \nu\frac{x^2}{(1+|x|^2)^2}~, \nonumber
\\[1.0ex]
    H^{\pm}_nx_n &=& C_1x_n+C_2x_{n\pm1}~. \nonumber
\end{eqnarray}

and with

\begin{equation}
    \nu=\frac{\alpha}{\kappa}\times10^{-3}=-2.
\end{equation}

As the relevant solutions (stationary dark solitons which possess a phase
jump across the density dip, as well as bubbles with no phase jump
across the dip) live against the backdrop of a constant density
background, it is relevant to examine the linearization properties
of such a background. This can be done by using the following form:
\begin{equation}
    a_n=\phi\exp(ikn)~,\qquad b_n=\phi\exp(ikn)~,
\end{equation}
with $k$ being the wavenumber of the background. In our case, we have restricted our considerations
to $k=0$ (unstaggered background) and $k=\pi$ (staggered background). Introducing the relation above into Eq. (\ref{eq:stat}) we get:
\begin{equation}
    \beta=-\nu\frac{\phi^2}{1+\phi^2}-\epsilon(k)~,
\end{equation}
where $\epsilon(k)=C_1+C_2\cos(k)$. It is important to note here that
in order that $\phi\in\mathbb{R}$, the condition  that needs
to  be fulfilled reads:
\begin{equation}
    \beta\in[-\epsilon(k),-\nu-\epsilon(k)]~,
\end{equation}
which, when $k=0$, will turn out (in our numerical results below) to
coincide with the existence interval for dark solitons with unstaggered
background. More generally,
that condition constitutes a necessary one for
the existence of dark / bubble solitons. In fact, as we will see below
bubble solutions only exist in a sub-interval of the allowable
parametric interval.

The dispersion relation of the linear excitations corresponds to the
continuous spectrum that will be identified in the linearization around
our solitary wave solutions. This relation can
be identified by decomposing the perturbations as
$\{P_n,Q_n,R_n,S_n\}=\{P,Q,R,S\}\exp(iqn)$ in Eq.~(\ref{eq:stat}) and deriving
the resulting condition:

\begin{equation}\label{eq:lineardisp}
    \omega_\pm^2(q)=\epsilon(k)[\epsilon(k)-2h(k)]+g^2(q)\pm2[h(k)-\epsilon(k)]g(q)
\end{equation}

with

\begin{eqnarray}
    h(k) &=& -\frac{[\beta+\epsilon(k)][\nu+\beta+\epsilon(k)]}{\nu}~, \\
    g^2(q) &=& C_1^2+C_2^2+2C_1C_2\cos(q)~.
\end{eqnarray}

It is easy to check that Eq. (\ref{eq:lineardisp}) predicts the existence, when restricting to positive eigenfrequencies, of two bands of real eigenvalues $\omega_\pm$ when $k=0$; besides, that equation also predicts for $k=\pi$ the existence of a band $\omega_-$ with real eigenfrequencies and another band $\omega_+$ with imaginary eigenfrequencies. This amounts to the modulational instability
of the background \cite{Stepic2006} which is detrimental for the survival of the staggered
states mentioned in the previous section.


\section{Numerical Observations}

We now turn to numerical observations, as a way of illustrating the
existence and potential stability of the localized states under
consideration above.

\begin{figure}[!ht]
\begin{center}
\begin{tabular}{cc}
    \includegraphics[width=\smallerfig]{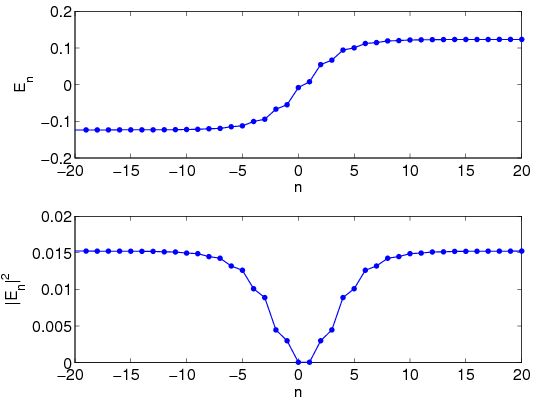} & \includegraphics[width=\smallerfig]{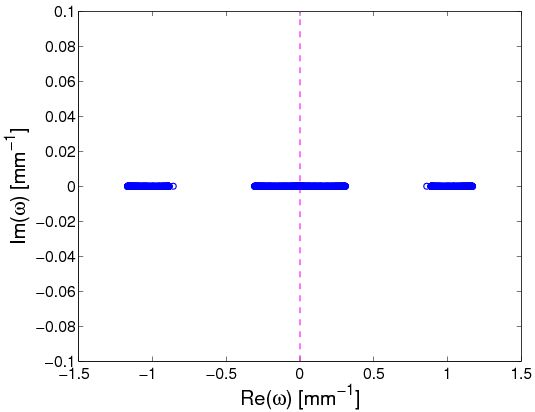} \\
\includegraphics[width=\smallerfig]{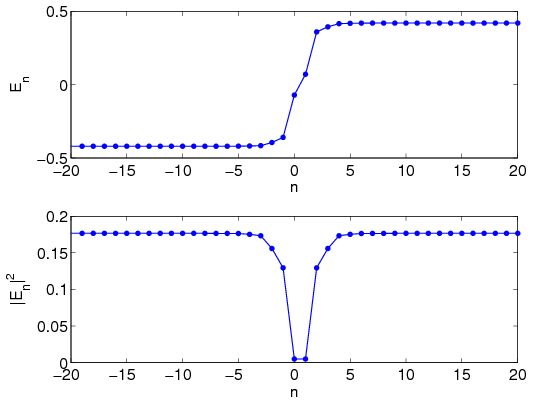} & \includegraphics[width=\smallerfig]{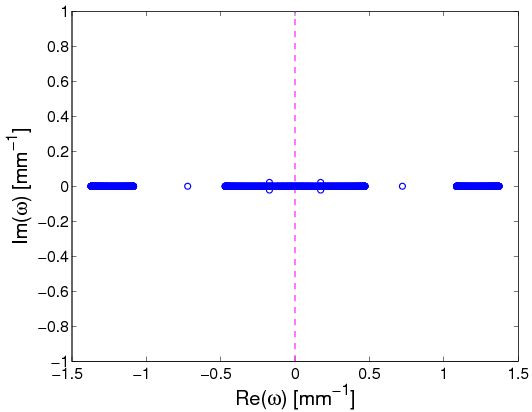} \\
\includegraphics[width=\smallerfig]{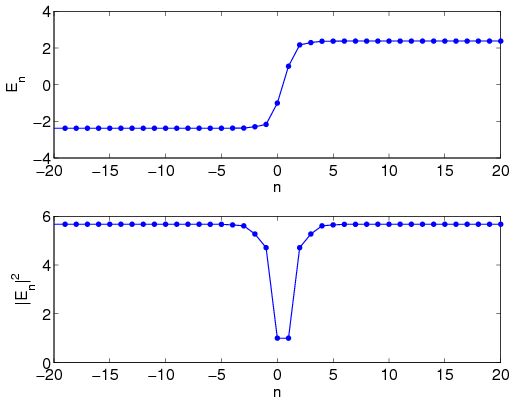} & \includegraphics[width=\smallerfig]{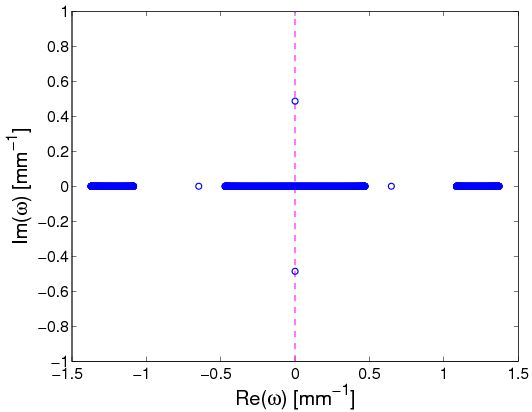}
\end{tabular}
\caption{(Left) Profile of the normalized electric field and the
corresponding intensity pattern for a dark soliton with $C_2=0.14$ and $\beta=-0.54$ (top), $\beta=-0.27$
(middle) and $\beta=1.13$ (bottom). The corresponding
right panels show the spectral stability plane. In the latter, the
existence of eigenfrequencies with non-vanishing imaginary part is
tantamount to instability with a growth rate equal to the absolute
value of this imaginary part.}
\label{fig:dark1}
\end{center}
\end{figure}

\begin{figure}[!ht]
\begin{center}
\begin{tabular}{cc}
    \includegraphics[width=\smallerfig]{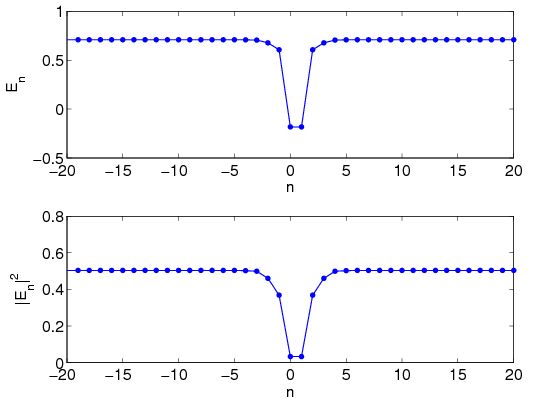} & \includegraphics[width=\smallerfig]{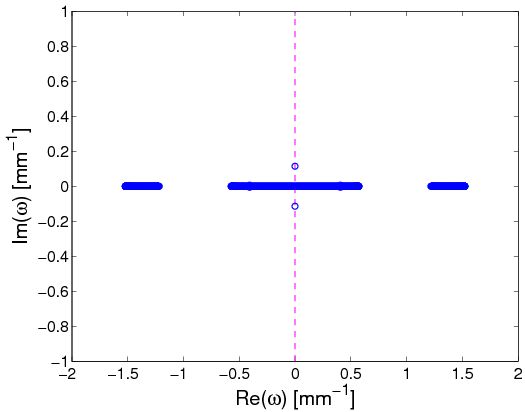} \\
\includegraphics[width=\smallerfig]{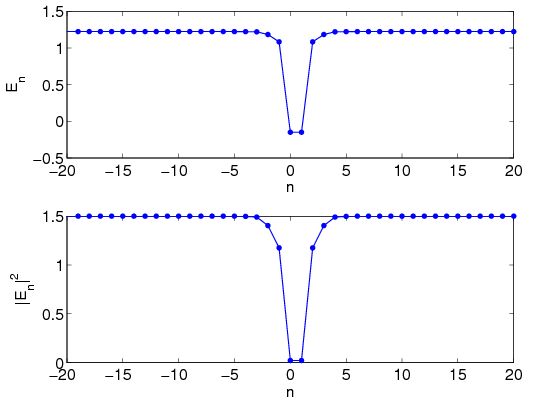} & \includegraphics[width=\smallerfig]{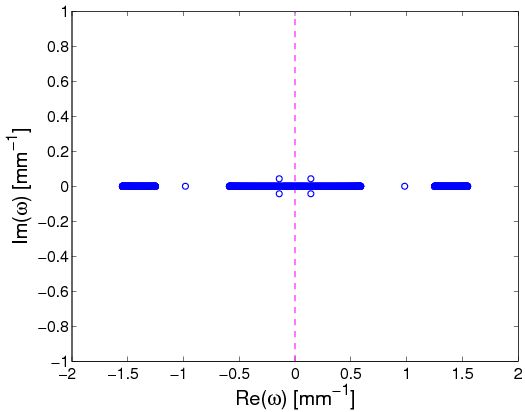} \\
    \includegraphics[width=\smallerfig]{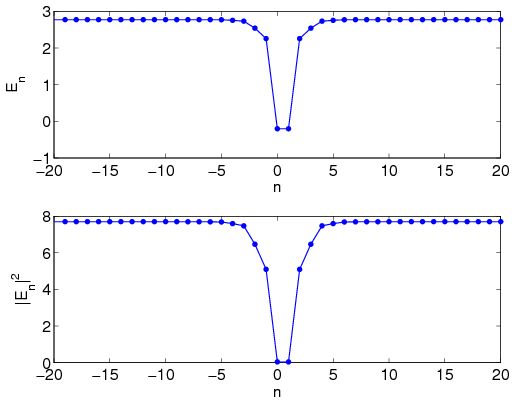} & \includegraphics[width=\smallerfig]{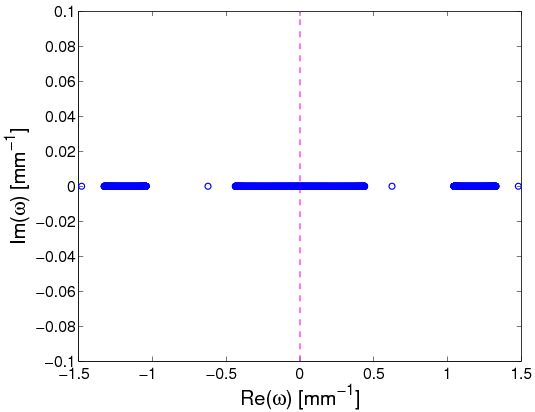}
\end{tabular}
\caption{(Left) Profile of the normalized electric field and the corresponding
intensity pattern for a bubble soliton with $C_2=0.14$ and $\beta=0.10$ (top
panel), $\beta=0.63$ (middle panel) as well as $\beta=1.2$ (bottom
panel). The corresponding right panels show the spectral stability plane.}
\label{fig:bubble1}
\end{center}
\end{figure}

\begin{figure}[ht]
\begin{center}
\begin{tabular}{cc}
    \multicolumn{2}{c}{Dark solitons. $C_2=0.14$} \\
    \includegraphics[width=\smallerfig]{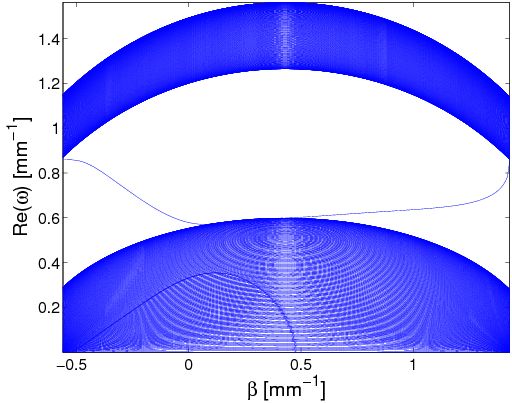} & \includegraphics[width=\smallerfig]{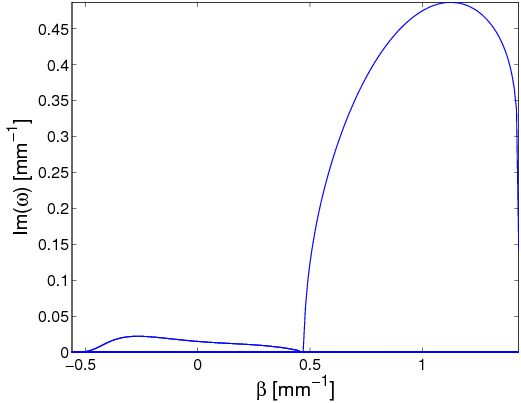} \\
    \multicolumn{2}{c}{Dark solitons. $C_2=0.05$} \\
    \includegraphics[width=\smallerfig]{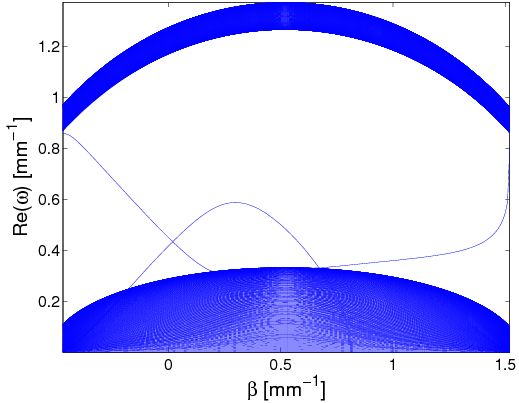} & \includegraphics[width=\smallerfig]{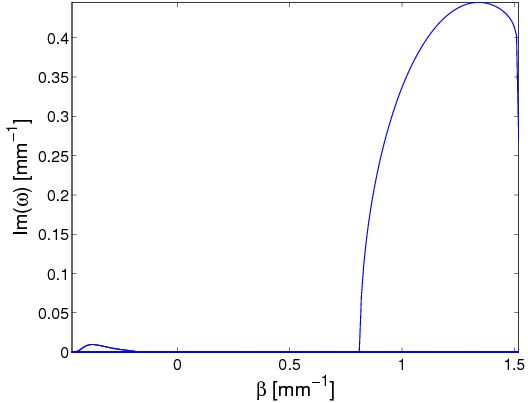} \\
    \multicolumn{2}{c}{Dark solitons. $C_2=0.50$} \\
    \includegraphics[width=\smallerfig]{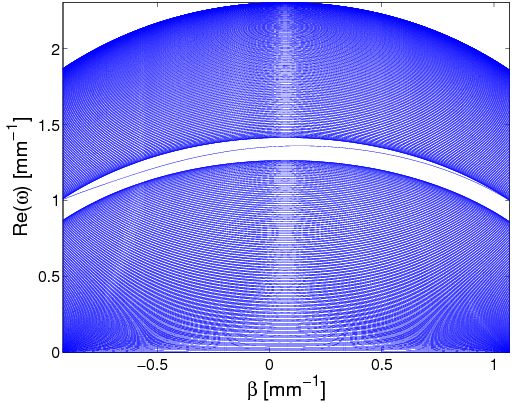} & \includegraphics[width=\smallerfig]{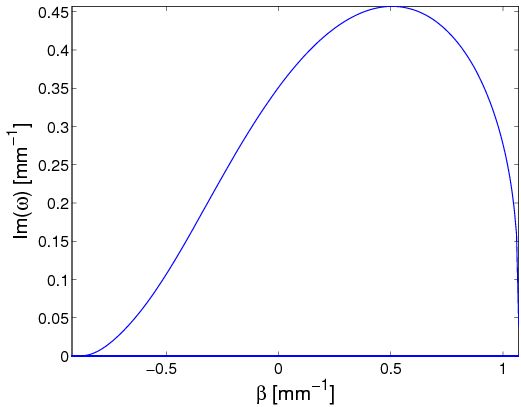} \\
    \multicolumn{2}{c}{Bubble solitons. $C_2=0.14$} \\
    \includegraphics[width=\smallerfig]{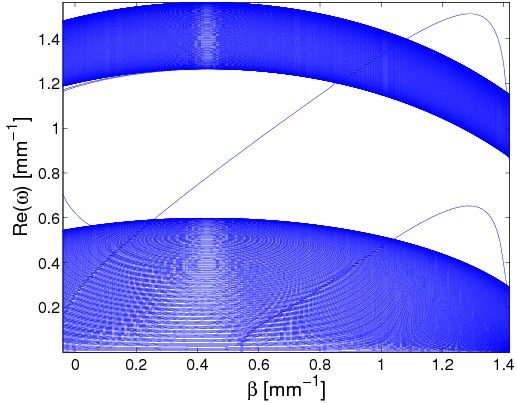} & \includegraphics[width=\smallerfig]{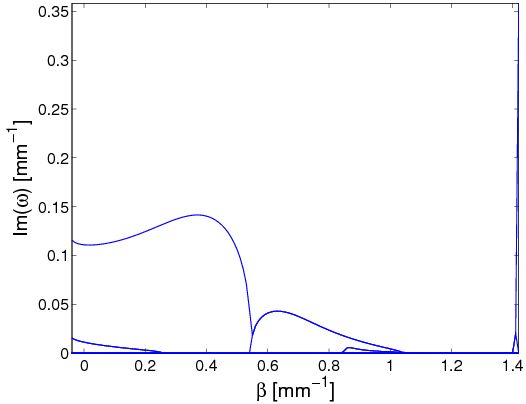}
\end{tabular}
\caption{Dependence with respect to $\beta$ of the real (left) and imaginary (right) part of the linearization eigenvalues for some selected values of $C_2$.}
\label{fig:stab1}
\end{center}
\end{figure}

Figs. \ref{fig:dark1}-\ref{fig:bubble1} show examples of the profile and the
spectral stability planes of dark and bubble solitons with
unstaggered backgrounds. It can be seen that as the propagation
parameter is varied, the discrete dark soliton solution may go from a regime
of spectral stability to one of weak oscillatory instability and finally
to one of stronger exponential instability. The former (oscillatory)
instability is associated with complex eigenfrequencies, while
the latter (exponential) instability with purely imaginary ones.
Similar features are observed for the bubbles presented in
Fig.~\ref{fig:bubble1}.
A fundamental difference appears to be that for dark solitons
an eigenfrequency pair associated with the wave first moves
along the real frequency axis. This leads to resonances
with the modes of the continuous spectrum, yielding
oscillatory instabilities until a point of maximal excursion i.e.,
maximal Re$(\omega)$. Thereafter, the real part of this eigenfrequency
starts decreasing and eventually crosses zero. Thus, it becomes
unstable as an imaginary eigenfrequency causing an exponential
instability. On the contrary, the relevant eigenfrequency pair first
becomes imaginary for bubbles, leading to an immediate exponential
instability. Then, upon a maximal excursion,
it returns to the origin, and subsequently moves along the real
frequency axis. This gives rise
to an oscillatory instability (although additional oscillatory
instabilities may also exist).
These features are presented for different values
of $C_2$ in Fig.~\ref{fig:stab1}. It is relevant to mention that
in that figure in addition to the experimentally relevant value
of $C_2=0.14$, presented both for dark solitons (top panel) and
for bubbles (bottom panel), cases of considerably smaller and
larger $C_2$ are examined in the middle panels. For smaller
$C_2$, the range of oscillatory instabilities shrinks (chiefly
because of the narrower interval of continuous spectrum and the
wider spectral gap in such a case). On the other hand, for larger
$C_2$, the  corresponding instability is exponential and is
found to arise throughout
the examined interval of propagation constants (see third panel
of Fig.~\ref{fig:stab1}). More specifically, for
$C_2=0.05$, the exponential instability in the right panel
of Fig.~\ref{fig:stab1}) only exists for the dark solitons with
$\beta \geq 0.81$; for $C_2=0.14$, it exists for
$\beta \geq 0.48$ and for $C_2=0.5$, it exists for
all $\beta \geq -0.92$. On the other hand, for the bubbles
of the bottom right panel of the figure, the exponential
instability is only present for $\beta \leq 0.54$
(notice the opposite sign of the inequality as discussed above).

\begin{figure}[ht]
\begin{center}
    \includegraphics[width=\middlefig]{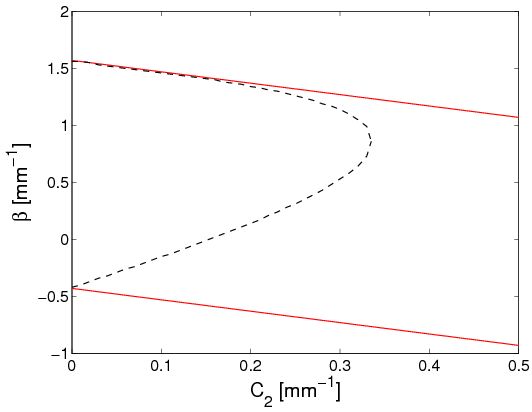}
\caption{Existence range of dark solitons and bubble solitons. Dark (bubble) solitons exist in the region between the full red (dashed black) lines.}
\label{fig:existence}
\end{center}
\end{figure}

\begin{figure}[ht]
\begin{center}
\begin{tabular}{cc}
    \includegraphics[width=\smallerfig]{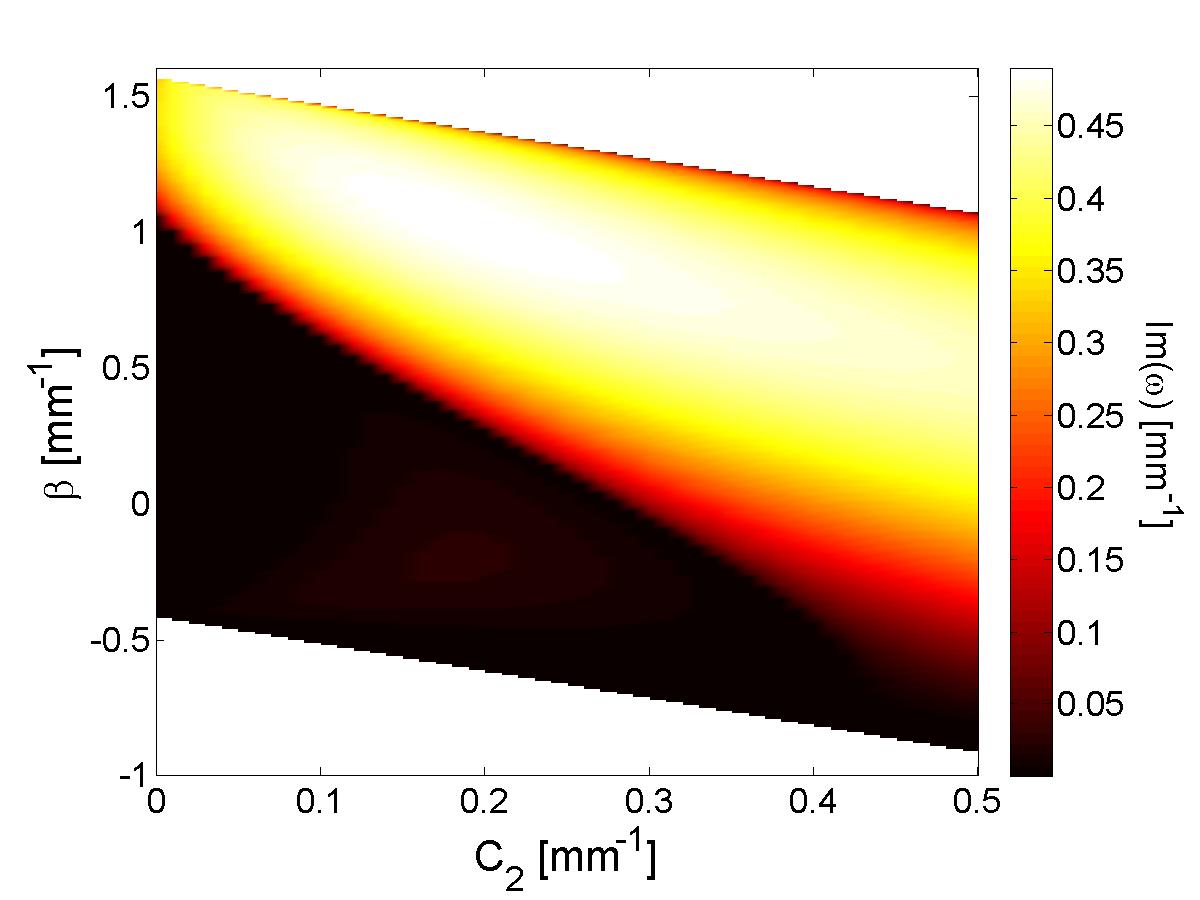} & \includegraphics[width=\smallerfig]{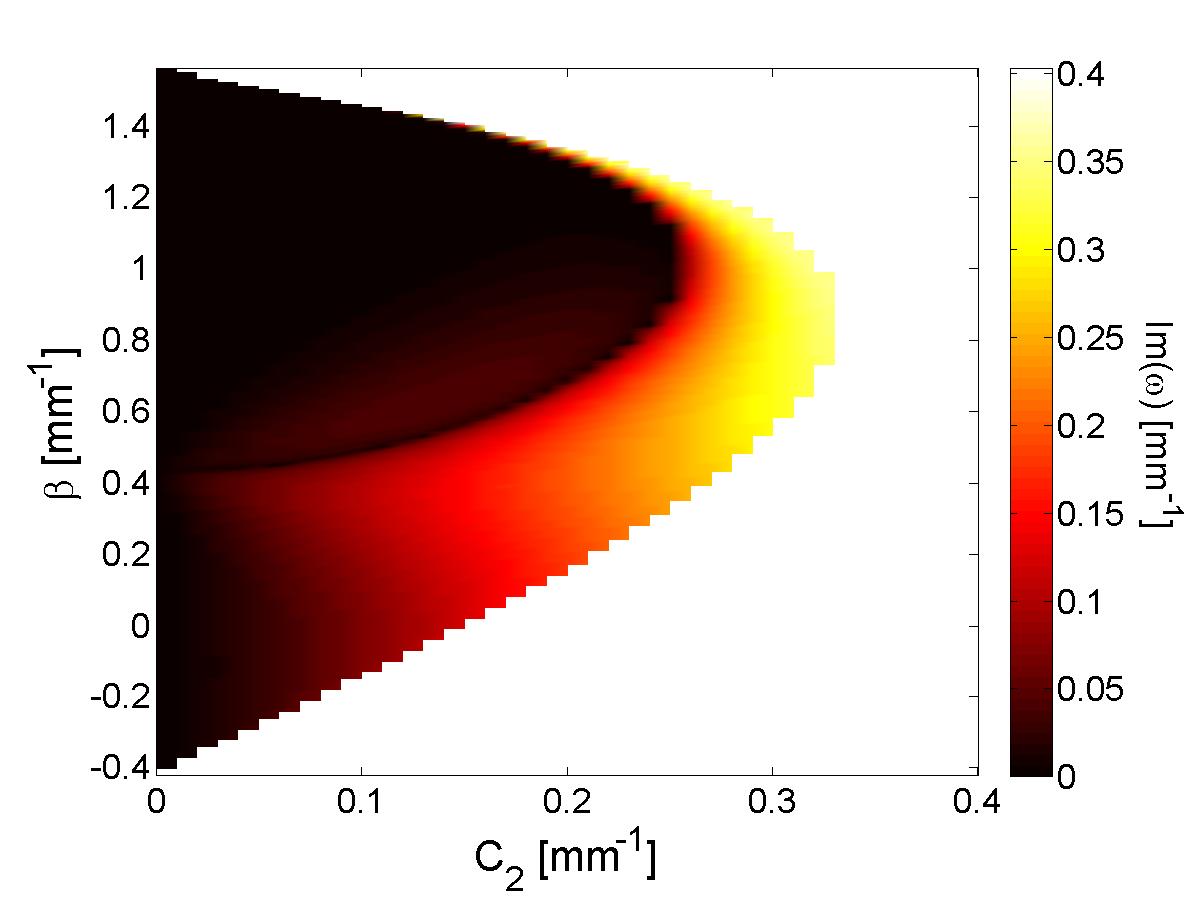} \\
    \includegraphics[width=\smallerfig]{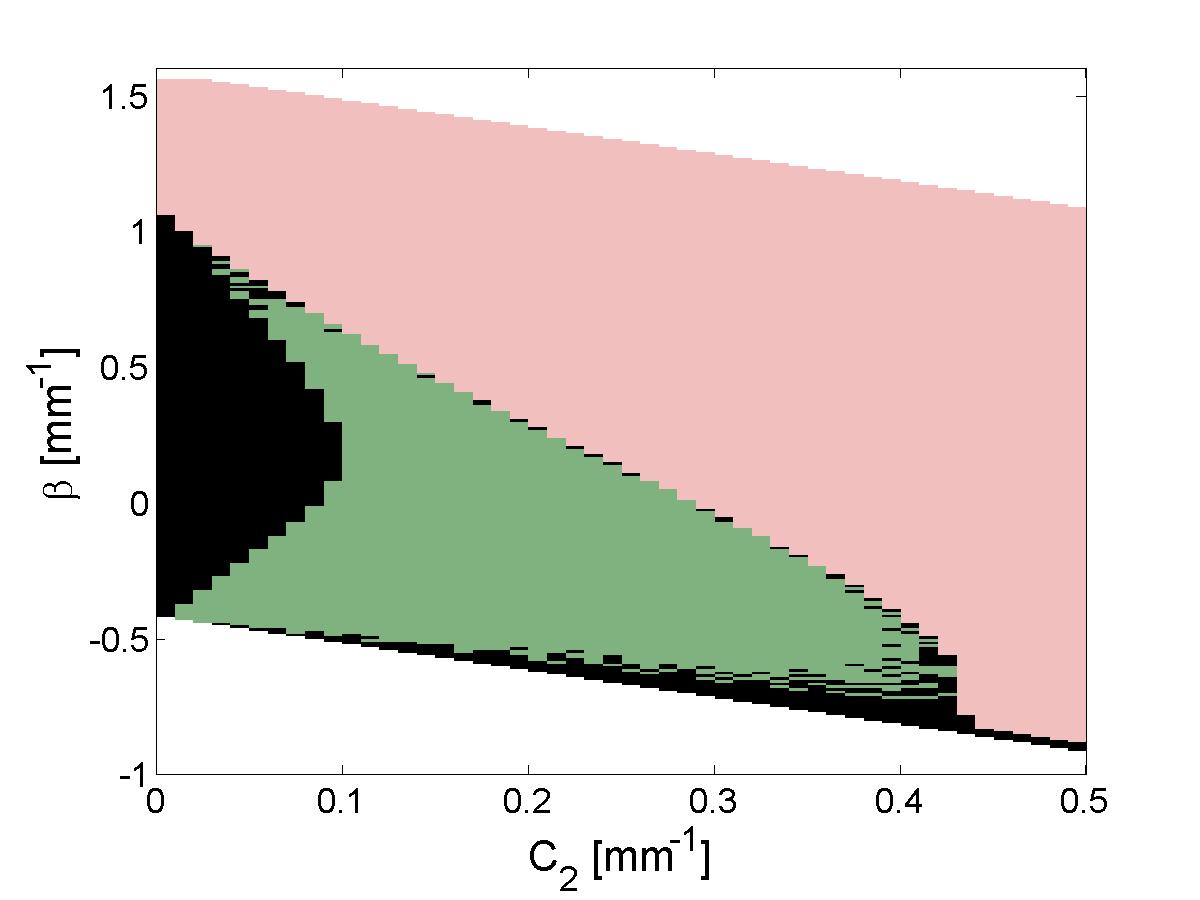} & \includegraphics[width=\smallerfig]{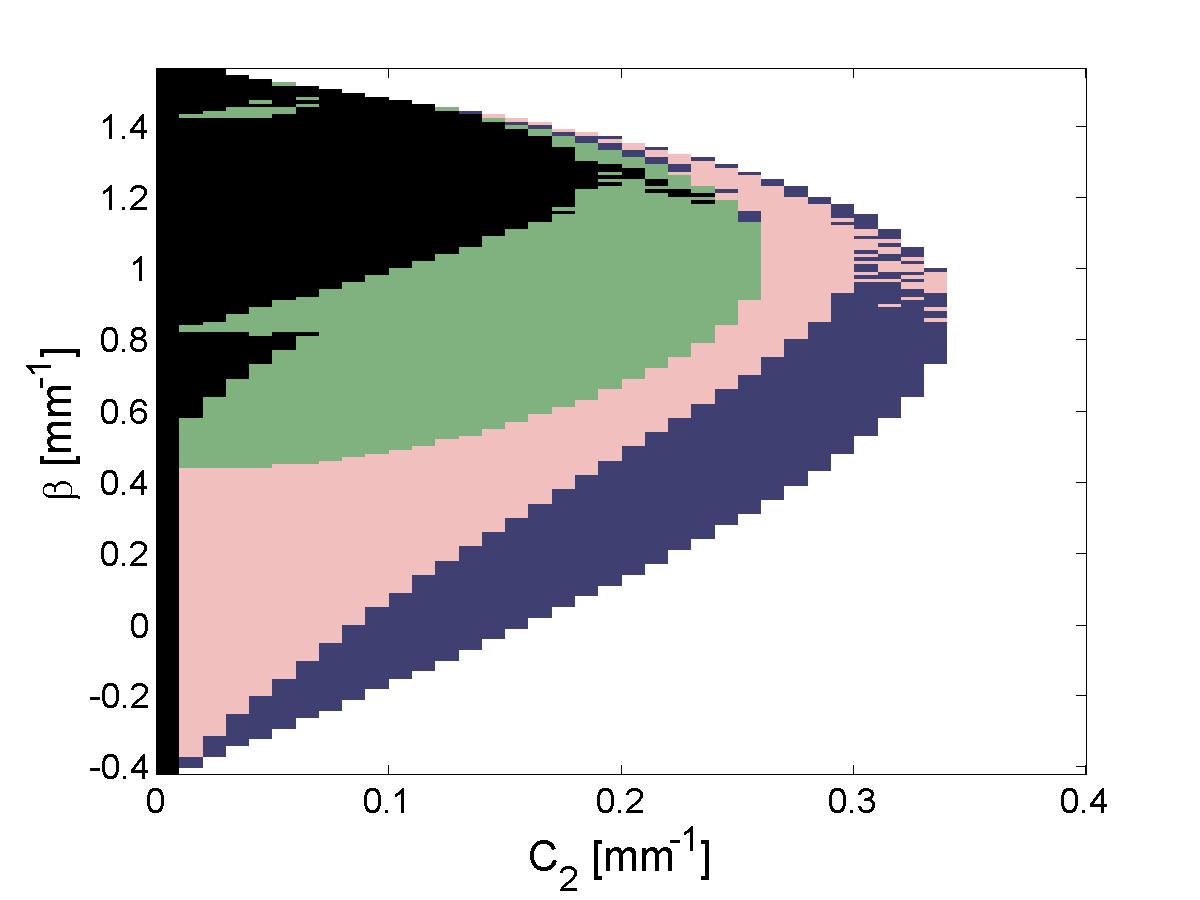} \\
\end{tabular}
\caption{(Top panels) Instability growth rates for dark (left) and bubble (right) solitons. The bottom panels show the type of instability experienced by the solitons, with the color coding as follows: black $\rightarrow$ stability;
green [medium gray] $\rightarrow$ oscillatory instability; pink [light gray] $\rightarrow$ exponential
instability; blue [dark gray] $\rightarrow$ oscillatory + exponential instabilities.}
\label{fig:stability}
\end{center}
\end{figure}

The detailed existence range of dark and bubble solitons is shown in
Fig.~\ref{fig:existence}. It is worth remarking that our numerics show that
dark solitons always exist in the range $\beta\in[-(C_1+C_2),-(C_1+C_2-\nu)]$;
however, the existence range of bubbles is limited (to a narrower
sub-interval),
and these solutions cease to exist for $C_2\geq0.34$. The stability properties
together with the relevant instability growth rates are summarized in
Fig.~\ref{fig:stability}. It should be noted that there exist
windows of stability
inside the region where the solitons are oscillatorily unstable; this fact
is caused by the finite lattice size (see e.g.,~\cite{johkiv} for a
relevant discussion) and is ubiquitous for dark solitons in lattices.
Furthermore, this phenomenon disappears in the limit where the
length of the lattice tends to infinity.

\begin{figure}[ht]
\begin{center}
\begin{tabular}{cc}
    \includegraphics[width=\smallerfig]{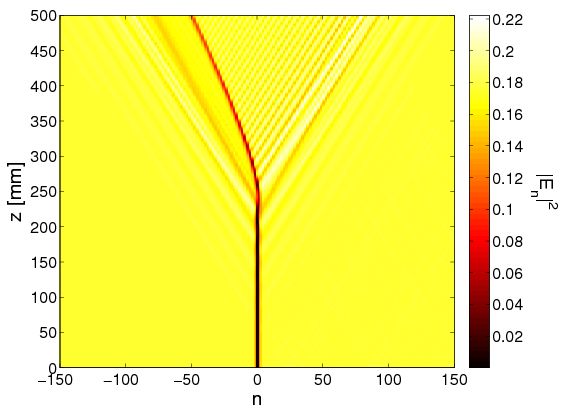} & \includegraphics[width=\smallerfig]{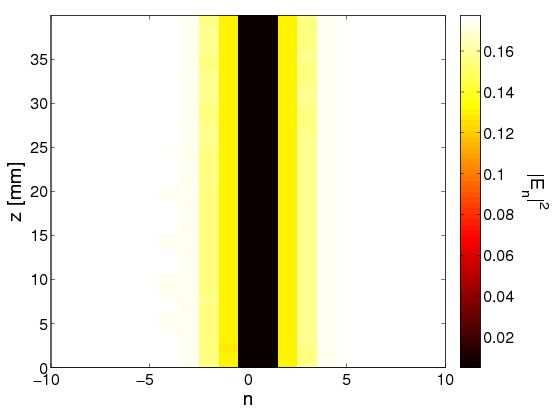} \\
\includegraphics[width=\smallerfig]{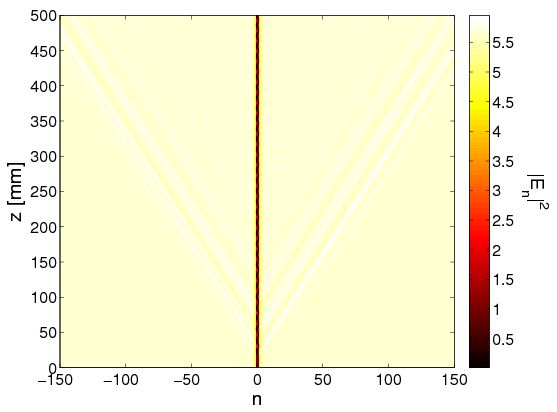} & \includegraphics[width=\smallerfig]{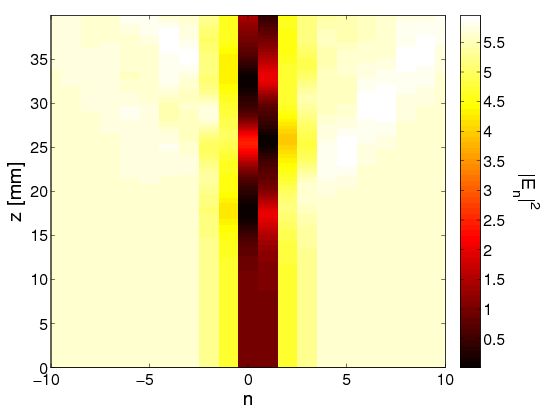}
\end{tabular}
\caption{Propagation of the unstable waves of the middle and bottom
panels of Fig. \ref{fig:dark1}, respectively, when a random perturbation of
amplitude $\sim10^{-3}$ is introduced. The right panel is a zoom of the left panel. In the left panel, the resolution of the propagation is $z=0.1$ mm, whereas in the right panel, the resolution is $z=0.01$ mm.}
\label{fig:propag1}
\end{center}
\end{figure}

\begin{figure}[ht]
\begin{center}
\begin{tabular}{cc}
    \includegraphics[width=\smallerfig]{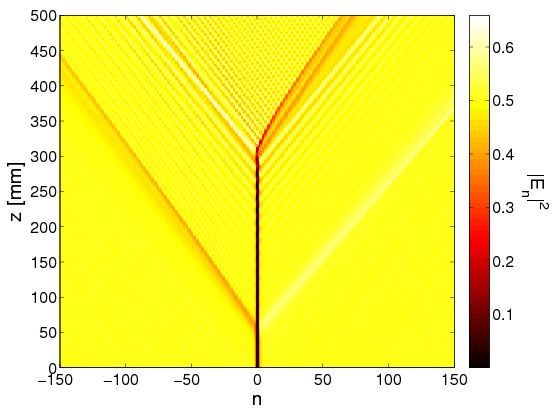} & \includegraphics[width=\smallerfig]{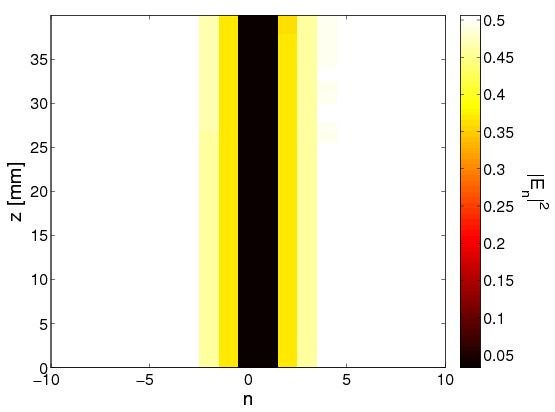} \\
 \includegraphics[width=\smallerfig]{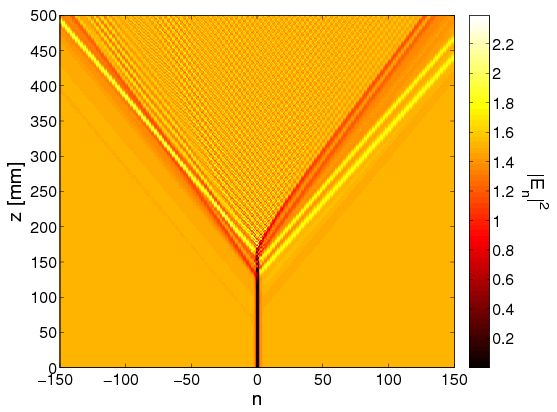} & \includegraphics[width=\smallerfig]{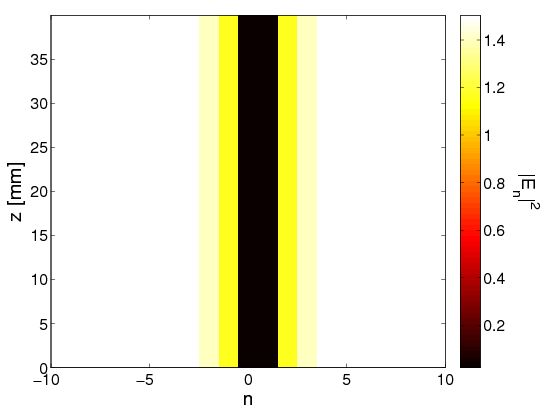}
\end{tabular}
\caption{Propagation of the unstable soliton of the top and
middle panels of Fig. \ref{fig:bubble1} when a random perturbation of
amplitude
$\sim10^{-3}$ is introduced. The right panel is a zoom of the left panel,
as before. In the left panel, the resolution of the propagation is
$z=0.1$ mm, whereas in the right panel, the resolution is $z=0.01$ mm.}
\label{fig:propag5}
\end{center}
\end{figure}

In Figs.~\ref{fig:propag1}-\ref{fig:propag5}, we show typical examples
of the dynamical evolution of the relevant waveforms, which have been obtained by a 4th order fixed-step Runge--Kutta method. In these examples,
we typically append to the unstable solutions a random perturbation of
amplitude $\sim10^{-3}$, so as to seed the instability (through the projection
of this random perturbation to the most unstable eigenmode of the solution).
The right panel of those figures is restricted to twice the sample length,
i.e., $z=40$ mm (in order to gauge the relevance of the reported phenomenology
to the experimental observations). Our general conclusions based on these
dynamical runs (as well as others not shown herein) is that
in most cases, the exponential instabilities may lead to a breathing of the
light intensity (i.e., $|E_n|^2$) or to a potential
destruction of the pertinent dynamical state.
An example of the former type is shown in the bottom panel
of Fig.~\ref{fig:propag1} and is justified by the fact that for
different values of the propagation constant $\beta$ (and all
other system parameters being identical), there exist other
members of this family of solutions which are, in fact, dynamically
stable. An example of the latter type is shown in the top
panel of Fig.~\ref{fig:propag5} for the pertinent bubble state.
On the other hand, instabilities associated with complex
quartets of eigenfrequencies are found to give rise
to an oscillatory growth that is eventually
seen to typically lead to propagation of excitations through
the lattice with a non-zero angle (i.e., moving solitary waves).
An example of this type can be seen e.g. for a dark soliton
in the top panel of  Fig.~\ref{fig:propag1}, while for
a bubble a similar observation but leading to a more
complex dynamical state can be seen in the bottom panel of
Fig.~\ref{fig:propag5}. Similar oscillatory instability
evolutionary outcomes for a homogeneous lattice can be found e.g.
in~\cite{chuby,fitrakis}.

As concerns the bubbles, we should note that their dynamical
instability in the cubic-quintic problem was systematically studied
in the works of~\cite{ibaras1,ibaras2} through a series of
analytical and numerical tools. There, it was found that
stationary bubbles were unstable through a splitting towards
stable traveling ones. The fundamental difference encountered here
e.g. in  Fig.~\ref{fig:propag5} with respect to that setting is
that in the present work, the state of vanishing intensity is
dynamically unstable and hence the instability is more likely
to break up the bubble into shallower propagating excitations,
rather than to an expanding nucleus of vanishing intensity.
This is what is observed through the asymmetric process
(seeded by random noise) in Fig.~\ref{fig:propag5}, although it
should also be added that in the presence of discreteness, the
existence of the so-called Peierls-Nabarro barrier renders it
unlikely that these excitations will maintain their
speed due to the emission of small amplitude
radiative wavepackets, as discussed e.g. in~\cite{ibaras3}.

\section{Conclusions}

In the present work, motivated by the experimental examination
of binary lattices with alternating couplings between two
distinct values $C_1$ and $C_2$, we have proceeded to theoretically
model and subsequently analyze such states in the context of optical
waveguide arrays. The two prototypical states under consideration
have been discrete dark solitons and bubble solutions. The distinctive
feature between these two waveforms is the existence (for dark-solitons)
or non-existence (for bubbles) of a phase jump across the density
dip associated with the solution. Fixing one of the two couplings,
we have chosen to vary the other coupling, as well as the propagation
constant of the solution (as a measure of the nonlinearity) and
to examine the existence and stability properties of such states.
We have found that dark solitons exist throughout the interval
(of couplings and/or of propagation constant) in which the background state
is found to exist. On the other hand, in the case of bubbles, the
range of existence is found to be somewhat more limited, a feature
that may be responsible for the experimental observation of their
eventual ``conversion'' to dark soliton states. The existence
of two tunable coupling parameters (or of the tunability of those in
conjunction with the propagation constant) is found to offer some
novel properties to the present model in connection to its standard
homogeneous dynamical lattice
counterpart. Firstly, the linear spectrum consists of two
bands with a ``mini-gap'' between them, which is controlled by these
parameters. This enables the existence of additional stationary
(staggered) states whose dynamical
instability has been briefly touched upon herein.
Furthermore, the parameters enable transitions between
oscillatory instabilities, dynamical stability
and exponential instabilities for the same type of state, a wealth of
possibilities that is not accessible in the standard DNLS model case.

It would be particularly interesting to extend the present considerations
to higher dimensional settings and especially to two-dimensional
arrays where different tunabilities of the spacings may be accessible.
On the one hand, there exists the tunability between different lattice
directions, essentially amounting to anisotropic nonlinear dynamical
lattices (see e.g. the early examinations thereof in~\cite{aniso}, where
interesting phenomena emerged from the breaking of the perfect
symmetry of the square lattice). Yet on the other hand, there is the
possibility of combining this anisotropy with binary couplings,
creating possibilities for inter- and intra-directional heterogeneity
and their interplay. In that regard, the effect of such heterogeneous
nonlinear dynamical
lattices on fundamental higher dimensional excitations such as discrete
vortices would be particularly interesting to explore in future work.

\begin{acknowledgement}

DK thanks the German Research Foundation (DFG, grant KI 482/11-2)
and the German-Israeli Foundation (GIF) for financial support of this research. PGK gratefully acknowledges support from the US-NSF through grants DMS-0806762 and CMMI-1000337 and from the Alexander von Humboldt Foundation as well as the Alexander
S. Onassis Public Benefit Foundation. JC acknowledges financial support from the MICINN project FIS2008-04848.

\end{acknowledgement}

\end{document}